\documentclass[aps,physrev,reprint,floatfix,groupedaddress,amsmath,amssymb,
longbibliography]{revtex4-2}

\pdftrailerid{}
\usepackage[T1]{fontenc} \usepackage{mathtools}
\usepackage{graphicx}
\usepackage{lineno}
\usepackage{physics}
\usepackage{mathrsfs}
\usepackage{bm}
\usepackage{float}

\usepackage{hyperref}
\usepackage[section]{placeins}
\usepackage{longtable}
\usepackage{multirow}
\usepackage{booktabs}
\usepackage{makecell}
\usepackage{xcolor}
\usepackage{dcolumn}
\usepackage[version=4]{mhchem}

\makeatletter
\newcommand{\warninput}[1]{\filename@parse{#1}\InputIfFileExists{#1}{}{\message{LaTeX Warning: File `\filename@base.\ifx\filename@ext\relax tex\else\filename@ext\fi' not found on input line \the\inputlineno}}}
\makeatother

\makeatletter
\begin{document}
    \title{Framework for atom-resolved decomposition of second-harmonic generation in nonlinear optical crystals}

    \author{YingXing Cheng}
\author{Congwei Xie}
    \author{Zhihua Yang}
    \email{zhyang@ms.xjb.ac.cn}
    \author{Shilie Pan}
    \email{slpan@ms.xjb.ac.cn}
    \affiliation{Research Center for Crystal Materials; CAS Key Laboratory of Functional Materials and Devices for Special Environmental Conditions; Xinjiang Key Laboratory of Functional Crystal Materials; Xinjiang Technical Institute of Physics and Chemistry, Chinese Academy of Sciences, 40-1 South Beijing Road, Urumqi 830011, China.}
    \affiliation{Center of Materials Science and Optoelectronics Engineering, University of Chinese Academy of Sciences, Beijing 100049, China.}

    \date{\today}

    \begin{abstract}
        In this work, we develop a new framework for computing atom-resolved contributions to optical properties based on atoms-in-molecules (AIM) schemes.
The formalism is independent of the specific AIM method and is made rigorous by partitioning momentum matrix elements into atomic contributions while exactly satisfying the relevant sum rules.
Zincblende GaP provides a two-atom demonstration of the atom-resolved second-harmonic-generation (SHG) analysis using five AIM partitioning methods.
The two-center Ga-P channels account for $80.6\%$--$82.2\%$ of the SHG response, despite the pronounced partition dependence of the atomic charges.
We then apply the framework to six representative ultraviolet nonlinear-optical crystals: $\beta$-\ce{BaB2O4} (BBO), \ce{LiB3O5} (LBO), \ce{CsB3O5} (CBO), \ce{CsLiB6O10} (CLBO), \ce{KBe2BO3F2} (KBBF), and \ce{LiCs2PO4} (LCPO).
The atom-triplet decomposition shows that two-center channels dominate the principal SHG components, while on-site terms are smaller and multicenter terms provide an important secondary contribution.
Although the AIM charges, detailed local percentages, and exact pathway ranks depend on the partition, the physical interpretation remains stable across the tested AIM methods.
B/O-only pathways dominate the leading responses of KBBF and LBO.
Mixed pathways involving B/O or P/O atoms and cations make substantial contributions to the leading responses of BBO, CBO, CLBO, and LCPO.
The decomposition framework therefore establishes a quantitative connection between local response pathways and the bulk nonlinear-optical response.
     \end{abstract}

    \maketitle
    \newpage

    \section{Introduction}
    \label{sec:introduction}
    Nonlinear-optical (NLO) crystals are essential for frequency conversion and optical-field control in a wide range of photonic applications.
Among the associated phenomena, second-harmonic generation (SHG) is particularly important because it provides an efficient route to coherent radiation at shorter wavelengths and underlies many practical laser-conversion technologies.\cite{Mutailipu2020}
For a crystal to be useful in SHG applications, it must combine a sufficiently large nonlinear response with appropriate optical transparency, birefringence, and material stability.
Understanding the microscopic origin of SHG is therefore a central problem in the design of new NLO materials.

Accurate calculation and mechanistic interpretation of SHG coefficients are essential for establishing structure--property relationships in NLO crystals.
Early analyses often relied on approximate models such as anion-group theory, in which the macroscopic SHG response is treated as a superposition of contributions from constituent structural groups.\cite{Chen2005}
In practice, early estimates of group SHG coefficients frequently employed semiempirical electronic-structure methods, including the Complete Neglect of Differential Overlap (CNDO) approximation.\cite{Chen2005}
With advances in quantum chemistry, \emph{ab initio} calculations for molecular units and structural motifs became feasible.
Software packages such as GAUSSIAN can provide accurate electronic-structure data for isolated groups, thereby strengthening group-based interpretations.\cite{Lin2001,g16}
Such approaches have been successfully applied to selected borate NLO crystals.\cite{Lin2001,Chen2005}

Over the past two decades, progress in density functional theory (DFT) and perturbation theory has enabled direct first-principles calculations of SHG in periodic solids, notably through the sum-over-states (SOS) formalism\cite{Ghahramani1991,Sipe1993,Lin1999} and density-functional perturbation theory (DFPT).\cite{Veithen2005}
Modern polarization theory also provides complementary formulations based on Berry phases\cite{King-Smith1993,Vanderbilt1993,Xiao2010} and Wannier functions,\cite{DalCorso1994,Wang2017,Ibanez-Azpiroz2018} which can improve numerical robustness and conceptual clarity in many contexts.
Despite these advances, extracting atom-specific information from such calculations remains nontrivial.
Standard first-principles workflows usually yield only the total SHG tensor.
A systematic local decomposition is generally unavailable, especially one that preserves the macroscopic response exactly.

Several strategies have been proposed to access atom-level information in SHG.
Historically, early attempts were formulated directly in real space.
A representative example is the real-space atom-cutting scheme (RSAC).\cite{Lin1999,Duan1999}
In RSAC, an atomic region is defined as an atom-centered sphere with a prescribed radius, and a motif region is taken as the union of the atomic regions belonging to that motif.
The electronic wave functions within a selected atomic or motif region are removed, and the modified wave functions are then used to evaluate the momentum-matrix elements entering the SOS expression for SHG.\cite{Duan1999}
The change in the SHG coefficients between calculations without and with atom or motif cutting is then attributed to the removed atom or motif.
While conceptually straightforward, RSAC partitions are usually controlled by empirical radii or geometric constructions and can become unreliable in strongly covalent environments or for atoms with markedly nonspherical charge distributions.
Moreover, practical implementations can introduce overlapping regions or ambiguous boundaries, which complicates a mathematically controlled attribution.\cite{Duan1999}

More recently, orbital-based attribution schemes have been developed, including atom response theory (ART)\cite{Cheng2018a} and Wannier-projection (WP) approaches.\cite{Lei2020}
These methods evaluate the response in a chosen orbital representation and then assign it to atoms through projection onto atomiclike orbitals, for example tight-binding atomic orbitals in ART\cite{Cheng2018a} and maximally localized Wannier functions in WP methods.\cite{Lei2020}
In their standard implementations, however, ART and WP do not explicitly resolve off-site multicenter terms and may therefore underrepresent contributions that are spatially delocalized beyond a single atomic center.
Such terms can be important because SHG is generally mediated by transition pathways with both intra-atomic and interatomic character.
Motivated by this consideration, recent studies have employed atomic-orbital-projection (AOP) techniques to analyze key dipole moments along SHG pathways and to examine how the inferred atomic contributions evolve with incident photon energy.\cite{Li2022e,Li2022a}
A further practical limitation is that ART and AOP implementations based on localized orbitals or Wigner--Seitz constructions, as commonly used for projected densities of states, typically require a truncation radius, which can lead to incomplete allocation of the electronic weight.

A more rigorous alternative is provided by atoms-in-molecules (AIM) schemes, in which continuous weight functions $w_A(\mathbf{r})\in[0,1]$ form a partition of unity, $\sum_A w_A(\mathbf{r})=1$, and thereby partition real space continuously among the atoms.
Incorporating such AIM weights into practical SHG analysis, however, is nontrivial within standard SOS-based workflows.
Recently, Chi proposed an atomic-space-tessellation (AST) method, in which AIM weights are introduced into an SHG-weighted electron-density analysis by using $w_A(\mathbf{r})$ as an additional weight function and then integrating over all space to extract atomic and off-site contributions.\cite{Chi2025}
Despite its advantages, the attribution of SHG to individual bands is not unique.

Although these approaches provide complementary physical pictures, a quantitative AIM decomposition of the momentum matrix elements entering SHG is still lacking.
Here we construct such a framework within periodic projector-augmented-wave (PAW) theory.
The atomic momentum matrices are Hermitian and exactly additive, so their SHG contributions recover the conventional macroscopic tensor without a residual term.

The interpretation uses three physical classes.
On-site terms contain one atomic label, two-center terms couple two atoms, and multicenter terms involve three distinct atoms.
Ordered atom triplets are retained only as the bookkeeping needed for the SHG contraction.
For analysis, their permutations are collected into unordered atomic or chemical-motif channels.
This provides signed local contributions that cannot be obtained from qualitative orbital plots alone.
Although the formalism also applies to linear and higher-order optical responses, the present study focuses on SHG.

The remainder of this paper is organized as follows.
Section~\ref{sec:methods} derives exactly additive atom-resolved momentum matrices within periodic PAW theory.
It then formulates their use for linear, second-order, and third-order optical responses, with SHG as the present application.
Section~\ref{sec:details} describes the electronic-structure calculations, five AIM partitioning methods, and SHG settings.
Section~\ref{sec:results} first uses two-atom GaP to demonstrate the complete atom-resolved decomposition.
It then applies the framework to six ultraviolet NLO crystals, compares partitioning methods, and analyzes atom- and element-triplet SHG contributions.
The section also verifies finite-frequency additivity for BBO at 1064~nm and examines numerical convergence.
Finally, Sec.~\ref{sec:summary} gives the main conclusions.
 
    \section{Methods}
    \label{sec:methods}
    The construction has one reusable step: decompose each momentum matrix element into Hermitian atomic terms that sum to the conventional matrix element.
We first define this decomposition in periodic PAW, then show briefly how the same terms enter linear and general nonlinear response, and finally specialize the formalism to SHG.
The response reconstructed by summing every atomic assignment is exact for a fixed AIM partition, whereas the value assigned to an individual atom is partition dependent.

\subsection{Atom-resolved momentum matrix elements in periodic PAW}
\label{sec:atom_p_matrix}

We consider two Bloch eigenstates at the same $\mathbf{k}$ point, $|i\rangle \equiv |n\mathbf{k}\rangle$ and $|j\rangle \equiv |m\mathbf{k}\rangle$.
Our goal is to construct an atom-resolved decomposition of the momentum matrix element $\langle i|\hat{\mathbf p}|j\rangle$ that is Hermitian, exactly additive over atoms, and compatible with periodic PAW theory.

Let $\{w_A(\mathbf r)\}$ be a set of nonnegative, cell-periodic weight functions forming a partition of unity,
\begin{align}
  w_A(\mathbf r)\ge 0,
  \qquad
  \sum_A w_A(\mathbf r)=1
  \quad \text{for all } \mathbf r \in \Omega ,
  \label{eq:aim_weights}
\end{align}
where $\Omega$ denotes the unit cell.
We associate with each weight function the corresponding multiplicative operator
\begin{align}
  \hat w_A \equiv w_A(\hat{\mathbf r})
  = \int_\Omega d^3r\, w_A(\mathbf r)\, |\mathbf r\rangle\langle \mathbf r| .
  \label{eq:wA_operator}
\end{align}
Because $w_A(\mathbf r)$ is real, $\hat w_A$ is Hermitian, $\hat w_A^\dagger=\hat w_A$, and Eq.~\eqref{eq:aim_weights} implies
\begin{align}
  \sum_A \hat w_A = \hat I .
  \label{eq:wA_sumrule}
\end{align}
In practice, $w_A(\mathbf r)$ may be obtained from smooth AIM schemes such as Hirshfeld and Hirshfeld-I,\cite{Hirshfeld1977,Bultinck2007} iterative stockholder analysis,\cite{C.Lillestolen2008} minimum-basis iterative stockholder analysis (MBIS),\cite{Verstraelen2016} linear approximation of iterative stockholder analysis (LISA),\cite{Benda2022,Cheng2025b,Cheng2025c} and additive variational Hirshfeld (AVH) partitioning.\cite{Heidar-Zadeh2024}
The present implementation requires smooth weights on the uniform fast Fourier transform (FFT) grid.
A discontinuous real-space partition such as a Bader basin\cite{Bader1972} is not directly interchangeable with these weights in the symmetrized momentum operator: its distributional boundary derivatives require a separate surface-term treatment.

To preserve Hermiticity at the operator level, we define the atomic contribution to the momentum operator by symmetric partitioning,
\begin{align}
  \hat{\mathbf p}_A
  \equiv
  \frac{1}{2}\left(\hat w_A \hat{\mathbf p} + \hat{\mathbf p}\hat w_A\right),
  \qquad
  \hat{\mathbf p}=-i\hbar\nabla .
  \label{eq:pA_def}
\end{align}
Since both $\hat w_A$ and $\hat{\mathbf p}$ are Hermitian, $\hat{\mathbf p}_A^\dagger=\hat{\mathbf p}_A$.
Moreover, using Eq.~\eqref{eq:wA_sumrule},
\begin{align}
  \sum_A \hat{\mathbf p}_A = \hat{\mathbf p},
\end{align}
so the decomposition is exactly additive.

Within PAW, matrix elements of a one-body operator separate into a smooth (pseudo) contribution and an on-site augmentation correction.
For the smooth part, we work with pseudo-wave functions $\tilde\psi_{n\mathbf k}(\mathbf r)$ represented on the real-space FFT grid.
We further define the masked pseudo-states
\begin{align}
  |\tilde i_A\rangle \equiv \hat w_A |\tilde i\rangle,
  \qquad
  |\tilde j_A\rangle \equiv \hat w_A |\tilde j\rangle ,
  \label{eq:masked_states}
\end{align}
with real-space representations
\begin{align}
  \tilde i_A(\mathbf r)=w_A(\mathbf r)\tilde\psi_i(\mathbf r),
  \qquad
  \tilde j_A(\mathbf r)=w_A(\mathbf r)\tilde\psi_j(\mathbf r).
\end{align}
The smooth atom-resolved momentum matrix element is then
\begin{align}
  \langle i|\hat{\mathbf p}_A|j\rangle^{\mathrm{smooth}}
  &= \frac{1}{2}\left(
    \langle \tilde i|\hat w_A \hat{\mathbf p}|\tilde j\rangle
    + \langle \tilde i|\hat{\mathbf p}\hat w_A|\tilde j\rangle\right)
  \nonumber \\
  &= \frac{1}{2}\left(
    \langle \tilde i_A|\hat{\mathbf p}|\tilde j\rangle
    + \langle \tilde i|\hat{\mathbf p}|\tilde j_A\rangle\right).
  \label{eq:sym_smooth}
\end{align}
Equation~\eqref{eq:sym_smooth} avoids explicit evaluation of $\nabla w_A$ and is therefore convenient for plane-wave/FFT implementations.

In reciprocal space, the two half terms in Eq.~\eqref{eq:sym_smooth} are evaluated as
\begin{align}
  \langle \tilde i_A|\hat{\mathbf p}|\tilde j\rangle
  &= \hbar \sum_{\mathbf G}
  \tilde i^{\,*}_{A,\mathbf G}\,(\mathbf k+\mathbf G)\,\tilde\psi_{j,\mathbf G},
  \\
  \langle \tilde i|\hat{\mathbf p}|\tilde j_A\rangle
  &= \hbar \sum_{\mathbf G}
  \tilde\psi^{*}_{i,\mathbf G}\,(\mathbf k+\mathbf G)\,\tilde j_{A,\mathbf G},
  \label{eq:fft_half}
\end{align}
where $\tilde i_{A,\mathbf G}$ and $\tilde j_{A,\mathbf G}$ are the FFT coefficients of $\tilde i_A(\mathbf r)$ and $\tilde j_A(\mathbf r)$, respectively, and $\mathbf G$ runs over reciprocal lattice vectors of the plane-wave grid.

Using $\hat{\mathbf p}=-i\hbar\nabla$, Eq.~\eqref{eq:sym_smooth} can also be written in terms of the paramagnetic transition current density for the orbital pair $(i,j)$,
\begin{align}
  \mathbf j^{\mathrm{para}}_{ij}(\mathbf r)
  =
  \frac{\hbar}{2m_e i}
  \Big(
    \tilde\psi_i^*(\mathbf r)\nabla\tilde\psi_j(\mathbf r)
    -
    [\nabla\tilde\psi_i^*(\mathbf r)]\,\tilde\psi_j(\mathbf r)
  \Big),
\end{align}
that is, the local current-density matrix element between states $|i\rangle$ and $|j\rangle$.
The smooth atom-resolved momentum matrix element is therefore
\begin{align}
  \langle i|\hat{\mathbf p}_A|j\rangle^{\mathrm{smooth}}
  =
  m_e \int_\Omega d^3r\,
  w_A(\mathbf r)\,\mathbf j^{\mathrm{para}}_{ij}(\mathbf r),
  \label{eq:current_form}
\end{align}
which makes the exact additivity with respect to the partition of unity explicit.

For a general one-body operator $\hat O$ (including $\hat{\mathbf p}$), the PAW augmentation correction is expressed in terms of partial waves $\{|\phi_\mu^A\rangle,|\tilde\phi_\mu^A\rangle\}$ and projectors $|\tilde p_\mu^A\rangle$ as
\begin{align}
  \Delta O^{(A)}_{\mu\nu}
  &\equiv \langle \phi_\mu^A|\hat O|\phi_\nu^A\rangle
  - \langle \tilde\phi_\mu^A|\hat O|\tilde\phi_\nu^A\rangle,
  \nonumber \\
  P_\mu^{(A)}(i)&\equiv \langle \tilde p_\mu^A|\tilde\psi_i\rangle .
\end{align}
The corresponding atom-resolved on-site term is
\begin{align}
  \langle i|\hat O|j\rangle_A^{\mathrm{aug}}
  =
  \sum_{\mu,\nu}
  P_{\mu}^{(A)*}(i)\,
  \Delta O^{(A)}_{\mu\nu}\,
  P_{\nu}^{(A)}(j).
  \label{eq:paw_aug}
\end{align}

In the implementation, $\hat O=\hat{\mathbf p}=-i\hbar\nabla$, and the PAW correction stored by \textsc{GPAW} is
\begin{align}
  \Delta \mathbf p^{(A)}_{\mu\nu}
  =
  -i\hbar\left[
  \langle\phi_\mu^A|\nabla|\phi_\nu^A\rangle
  -
  \langle\tilde\phi_\mu^A|\nabla|\tilde\phi_\nu^A\rangle
  \right].
  \label{eq:paw_momentum_correction}
\end{align}
Thus, the smooth contribution is partitioned using the AIM weights, whereas the complete on-site correction is assigned to the atom on which the augmentation sphere and projectors are centered.
No part of the PAW correction is discarded or multiplied by a real-space AIM weight.

Combining the smooth contribution, Eq.~\eqref{eq:sym_smooth}, with the on-site augmentation, Eq.~\eqref{eq:paw_aug}, we obtain the atom-resolved momentum matrix element
\begin{align}
  \langle i|\hat{\mathbf p}|j\rangle_A
  \equiv
  \langle i|\hat{\mathbf p}_A|j\rangle^{\mathrm{smooth}}
  +
  \sum_{\mu,\nu}
  P_{\mu}^{(A)*}(i)\,
  \Delta \mathbf p^{(A)}_{\mu\nu}\,
  P_{\nu}^{(A)}(j).
  \label{eq:final_atom_p}
\end{align}
Equivalently,
\begin{align}
  \langle i|\hat{\mathbf p}|j\rangle_A
  &= \frac{1}{2}\left(
    \langle \tilde i_A|\hat{\mathbf p}|\tilde j\rangle
    + \langle \tilde i|\hat{\mathbf p}|\tilde j_A\rangle\right)
  \nonumber \\
  &\quad + \sum_{\mu,\nu} P_{\mu}^{(A)*}(i)\,
  \Delta \mathbf p^{(A)}_{\mu\nu}\,P_{\nu}^{(A)}(j).
\end{align}
By construction, this decomposition is Hermitian,
\begin{align}
  \langle i|\hat{\mathbf p}|j\rangle_A
  =
  \langle j|\hat{\mathbf p}|i\rangle_A^{*},
\end{align}
and exactly additive,
\begin{align}
  \sum_A \langle i|\hat{\mathbf p}|j\rangle_A
  =
  \langle i|\hat{\mathbf p}|j\rangle,
\end{align}
provided that Eq.~\eqref{eq:aim_weights} holds.
While the numerical distribution among atoms depends on the choice of $\{\hat w_A\}$, the sum rule is exact.

\subsection{Linear optical response}
\label{subsec:linear_response}

We define the transition frequency as $\omega_{nm}\equiv\omega_n-\omega_m=(E_n-E_m)/\hbar$, where $E_n$ and $E_m$ are the energies of bands $n$ and $m$, respectively.
Throughout the linear and nonlinear response formulas below, causality is imposed by replacing $\omega$ with $\omega+i\eta$ and taking the limit $\eta\rightarrow0^+$.
For $n\neq m$, the interband Berry connection is
\begin{align}
  r^a_{nm}=i\langle u_n|\partial_{k_a}|u_m\rangle,
\end{align}
where $|u_n\rangle$ is the cell-periodic part of the Bloch state.
For distinct, nondegenerate bands, it can be written in terms of velocity matrix elements as
\begin{align}
  r_{nm}^a = -\,i\,\frac{v_{nm}^a}{\omega_{nm}}.
  \label{eq:r_p_relation}
\end{align}
When the local-Hamiltonian relation $\hat{\mathbf p}=m_e\hat{\mathbf v}$ applies, where $m_e$ is the free-electron mass, this becomes
\begin{align}
  r_{nm}^a = \frac{p_{nm}^a}{i\,m_e\omega_{nm}}.
  \label{eq:r_from_p}
\end{align}
Here, ``velocity'' refers to that associated with the reconstructed local all-electron Hamiltonian, rather than the bare canonical momentum of the nonlocal pseudo-Hamiltonian.
Equation~\eqref{eq:r_p_relation} is undefined within an exactly degenerate subspace, where the individual off-diagonal matrix elements depend on the chosen basis.
For exactly degenerate distinct bands, the wave functions may be chosen so that the relevant off-diagonal matrix elements vanish.\cite{Aversa1995,Rashkeev1998}
Numerically, we use the corresponding regularized convention and set $r_{nm}^a$ to zero when $|\omega_{nm}|$ is below the energy-difference tolerance used in the SHG calculation.

For a cold semiconductor, the independent-particle linear susceptibility is\cite{Aversa1995,Gajdos2006}
\begin{align}
  \chi^{(1)}_{ab}(\omega)
  &= \frac{e^2}{\hbar}
  \int_{\mathrm{BZ}}\frac{d\mathbf{k}}{(2\pi)^3}
  \sum_{n\in V,m\in C}
  \nonumber\\
  &\quad\times\left[
    \frac{r^a_{nm}r^b_{mn}}{\omega_{mn}-\omega}
    +\frac{r^b_{nm}r^a_{mn}}{\omega_{mn}+\omega}
  \right],
  \label{eq:linear_chi}
\end{align}
where $V$ and $C$ denote the fully occupied valence and empty conduction bands, respectively, and $e$ is the elementary charge.
Applying Eq.~\eqref{eq:r_from_p} to the atom-resolved momentum matrix elements defines
\begin{align}
  r_{nm,A}^a=\frac{p_{nm,A}^a}{i\,m_e\omega_{nm}},
  \qquad
  r_{nm}^a=\sum_A r_{nm,A}^a,
  \label{eq:atom_r_relation}
\end{align}
for distinct, nondegenerate bands.
Substitution of this sum in every Berry-connection factor gives
\begin{align}
  \chi^{(1)}_{ab}(\omega)
  =\sum_{A,B}\chi^{(1)}_{ab;AB}(\omega),
  \label{eq:linear_atom_pair}
\end{align}
where $\chi^{(1)}_{ab;AB}$ is obtained from Eq.~\eqref{eq:linear_chi} by replacing $r^a_{nm}r^b_{mn}$ with $r^a_{nm,A}r^b_{mn,B}$ and applying the analogous replacement to the second term.
Thus linear response contains two atomic labels: $A=B$ assigns both factors to one atom, whereas $A\ne B$ couples two atomic assignments.

\subsection{General nonlinear response and SHG}
\label{subsec:shg_from_p}

The length-gauge formalism of Aversa and Sipe provides well-behaved independent-particle expressions for second- and third-order susceptibilities at arbitrary frequency mixing.\cite{Aversa1995}
Each response is multilinear in interband matrix elements, intraband differences, and generalized derivatives constructed from the same electronic quantities.
Applying Eq.~\eqref{eq:final_atom_p} to every such factor gives the general sum rule
\begin{align}
  \chi^{(q)}=\sum_{A_0,\ldots,A_q}\chi^{(q)}_{A_0\ldots A_q},
  \qquad q=1,2,3,
  \label{eq:general_response_partition}
\end{align}
without changing the band energies, occupations, or frequency denominators.
Second-order responses therefore carry three atomic labels and third-order responses carry four.

We now specialize to SHG, $\chi^{abc}(-2\omega;\omega,\omega)$, using the length-gauge formulation of Rashkeev \textit{et al.}\cite{Rashkeev1998} For an independent-particle system with band occupations $f_n$, the second-order susceptibility is written as
\begin{align}
  \chi^{abc}(-2\omega;\omega,\omega)
  =
  \chi^{abc}_e(-2\omega;\omega,\omega)
  +
  \chi^{abc}_i(-2\omega;\omega,\omega),
\end{align}
where $\chi^{abc}_e$ and $\chi^{abc}_i$ denote the purely interband and mixed interband--intraband contributions, respectively.
Here $a$, $b$, and $c$ denote Cartesian directions.

The interband contribution reads\cite{Rashkeev1998}
\begin{align}
  \chi^{abc}_e(-2\omega;\omega,\omega)
  &=
  \frac{e^3}{\hbar^2}
  \int \frac{d\mathbf{k}}{(2\pi)^3}\sum_{nml}
  \frac{ r^a_{nm} \{r^b_{ml} r^c_{ln} \}}{\omega_{ln} - \omega_{ml}}
  \nonumber \\
  &\;\times
  \left[
      \frac{2f_{nm}}{\omega_{mn} - 2\omega}
    + \frac{f_{ln}}{\omega_{ln} - \omega}
    + \frac{f_{ml}}{\omega_{ml} - \omega}
  \right],
  \label{eq:chi_e_dyn}
\end{align}
with the symmetrized product
\begin{align}
  \{ r^b_{ml} r^c_{ln} \}
  =
  \frac{1}{2}\left( r^b_{ml} r^c_{ln} + r^c_{ml} r^b_{ln} \right).
\end{align}
The band indices are $n$, $m$, and $l$.
We define $f_{nm}\equiv f_n-f_m$ as the occupation difference.

The mixed interband--intraband contribution is\cite{Rashkeev1998}
\begin{widetext}
\begin{align}
  \chi_i^{abc}(-2\omega;\omega,\omega)
  &=
  \frac{i}{2}\,\frac{e^{3}}{\hbar^{2}}
  \int \frac{d \mathbf{k}}{(2\pi)^3}
  \sum_{n m} f_{nm}
  \Bigg[
  \frac{2}{\omega_{mn}\bigl(\omega_{mn}-2\omega\bigr)}\,
  r^{a}_{nm}\!\left(r^{b}_{mn;c}+r^{c}_{mn;b}\right)
  \nonumber \\
  &\quad+\frac{1}{\omega_{mn}\bigl(\omega_{mn}-\omega\bigr)}\,
  \left(r^{a}_{nm;c}\,r^{b}_{mn}+r^{a}_{nm;b}\,r^{c}_{mn}\right)
  \nonumber \\
  &\quad+\frac{1}{\omega_{mn}^{2}}
  \left(
  \frac{1}{\omega_{mn}-\omega}-\frac{4}{\omega_{mn}-2\omega}
  \right)
  r^{a}_{nm}\!\left(r^{b}_{mn}\Delta^{c}_{mn}+r^{c}_{mn}\Delta^{b}_{mn}\right)
  \nonumber \\
  &\quad -\frac{1}{2\,\omega_{mn}\bigl(\omega_{mn}-\omega\bigr)}
  \left(r^{b}_{nm;a}\,r^{c}_{mn}+r^{c}_{nm;a}\,r^{b}_{mn}\right)
  \Bigg],
  \label{eq:chi_i_dyn}
\end{align}
\end{widetext}
where $r^a_{nm;b}$ denotes the generalized derivative.\cite{Sipe1993,Sipe2000,Wang2024}
It is defined as
\begin{align}
  r^a_{nm;b}
  =
  \frac{\partial r^a_{nm}}{\partial k_b}
  - i\left(\xi^b_{nn}-\xi^b_{mm}\right) r^a_{nm},
\end{align}
with the intraband Berry connection $\xi^b_{nn}= i\langle u_n|\partial_{k_b}|u_n\rangle$.
The velocity-difference vector is $\Delta_{nm}^a = v_{nn}^a - v_{mm}^a$, where $v^a_{nn}$ is the band velocity.
For a local Hamiltonian, $\Delta_{mn}^a=(p_{mm}^a-p_{nn}^a)/m_e$.

The generalized derivative can be evaluated through the sum rule,\cite{Sipe1993,Wang2017}
\begin{align}
  r^b_{nm;a}
  &=
  \frac{r_{nm}^{a}\,\Delta_{mn}^{b}
        + r_{nm}^{b}\,\Delta_{mn}^{a}}{\omega_{nm}}
  \nonumber\\
  &\quad + \frac{i}{\omega_{nm}}
    \sum_{l}
    \bigl(\omega_{lm}\,r_{n l}^{a} r_{l m}^{b}
         - \omega_{n l}\,r_{n l}^{b} r_{l m}^{a}\bigr),
  \label{eq:gen_deriv_sum_rule}
\end{align}
which is exact for local Hamiltonians.
For nonlocal potentials, an additional term proportional to $\langle n|\partial_{k_b}\partial_{k_a}\hat{H}|m\rangle$ generally appears and may need to be retained.\cite{Wang2017,Cheng2026}

In practical calculations, a scissor correction is often applied to compensate for the band-gap underestimation of local and semilocal exchange--correlation functionals.
We consider two widely used protocols, scheme-L\cite{Levine1989,Levine1990,Levine1991} and scheme-N.\cite{Nastos2005}
Both protocols apply a rigid shift $\Delta>0$ to conduction-band energies in order to reproduce the experimental band gap, but they differ in how this shift is propagated into the SHG expressions and, in particular, into the generalized-derivative terms.

In the static limit, compact expressions with explicit Kleinman symmetry are available.
These formulas reorganize the response into three-band momentum products and should be distinguished from evaluating the separate length-gauge $\chi_e^{abc}(\omega)$ and $\chi_i^{abc}(\omega)$ expressions as $\omega\rightarrow0$.
We use the latter representation when examining convergence of the separated terms, retaining both contributions before taking their sum.
In our previous work, we derived numerically stable compact static-limit formulas for both scheme-L and scheme-N that avoid spurious divergences.\cite{Cheng2026}

In scheme-N,\cite{Nastos2005} the static-limit susceptibility can be written as
\begin{widetext}
\begin{align}
  \chi^{abc}
  &=
  \frac{e^{3}}{2\hbar^{2}}
  \int \frac{d \mathbf{k}}{(2\pi)^3}
  \sum_{n\in V}\sum_{m\in C}\sum_{l\in V}
  P(abc)\,\Im\!\left\{ p^{a}_{nm} p^{b}_{ml} p^{c}_{ln} \right\}
  \frac{1}{S_{mn}^2 \omega_{nm} \omega_{lm}}
  \left( \frac{1}{S_{lm}}
  + \frac{2}{\omega_{nm}} \right)
  \nonumber\\
  &\quad
  + \frac{e^{3}}{2\hbar^{2}}
  \int \frac{d \mathbf{k}}{(2\pi)^3}
  \sum_{n\in V}\sum_{m\in C}\sum_{l\in C}
  P(abc)\,\Im\!\left\{ p^{a}_{nm} p^{b}_{ml} p^{c}_{ln} \right\}
  \frac{1}{S_{mn}^2 \omega_{mn} \omega_{ln}}
  \left( \frac{1}{S_{ln}}
  + \frac{2}{\omega_{mn}} \right),
  \label{eq:schemeN_static}
\end{align}
\end{widetext}
where
\begin{align}
  S_{nm} = \omega_{nm} + f_{nm}\Delta,
\end{align}
and $\Delta>0$ is the scissor shift.
Here $V$ and $C$ denote valence and conduction bands, respectively.

In scheme-L,\cite{Levine1990} the corresponding static-limit susceptibility is
\begin{widetext}
\begin{align}
  \chi^{abc}
  &=
  \frac{e^{3}}{2\hbar^{2}}
  \int \frac{d \mathbf{k}}{(2\pi)^3}
  \sum_{n\in V}\sum_{m\in C}\sum_{l\in V}
  P(abc)\,\Im\!\left\{ \tilde{p}^{a}_{nm} \tilde{p}^{b}_{ml} \tilde{p}^{c}_{ln} \right\}
  \left(
      \frac{1}{S_{nm}^3 S_{lm}^2}
    + \frac{2}{S_{nm}^4 S_{lm}}
  \right)
  \nonumber\\
  &\quad
  + \frac{e^{3}}{2\hbar^{2}}
  \int \frac{d \mathbf{k}}{(2\pi)^3}
  \sum_{n\in V}\sum_{m\in C}\sum_{l\in C}
  P(abc)\,\Im\!\left\{ \tilde{p}^{a}_{nm} \tilde{p}^{b}_{ml} \tilde{p}^{c}_{ln} \right\}
  \left(
      \frac{1}{S_{mn}^3 S_{ln}^2}
    + \frac{2}{S_{mn}^4 S_{ln}}
  \right),
  \label{eq:schemeL_static}
\end{align}
\end{widetext}
where the scissor-renormalized momentum matrix elements are defined as
\begin{align}
  \tilde{p}^{a}_{nm}
  =
  \left[1 + \frac{\Delta}{\omega_{nm}}\left(\delta_{nC}-\delta_{mC}\right)\right] p^{a}_{nm}.
\end{align}
Here $\delta_{nC}=1$ if $n$ is a conduction band, and $\delta_{nC}=0$ otherwise, so that the renormalization applies only to valence--conduction transitions.

\subsection{Ordered atom-triplet-resolved SHG coefficients}
\label{subsec:atomic_shg}

The atom-resolved momentum matrix elements $p_{nm,A}^a$ are obtained from Eq.~\eqref{eq:final_atom_p}.
They satisfy the exact sum rule
\begin{align}
  p_{nm}^a=\sum_{A} p_{nm,A}^a,
  \qquad
  p_{nm,A}^a=\bigl(p_{mn,A}^a\bigr)^{*},
\end{align}
where $A$ labels atoms or, more generally, AIM fragments.
The corresponding atom-resolved Berry connections are given by Eq.~\eqref{eq:atom_r_relation}.
We define the atom-resolved velocity differences as
\begin{align}
  \Delta_{mn,A}^a = \frac{p_{mm,A}^a-p_{nn,A}^a}{m_e}.
\end{align}
Within the nondegenerate domain, these quantities inherit exact additivity,
\begin{align}
  r_{nm}^a=\sum_A r_{nm,A}^a,
  \qquad
  \Delta_{mn}^a=\sum_A \Delta_{mn,A}^a.
\end{align}
Consequently, any SHG expression that is multilinear in $r$ and $\Delta$ admits an exact decomposition into pathway contributions labeled by atom indices attached to the individual matrix-element factors.
We refer to these terms as \emph{ordered atom-triplet} contributions because the triplet $(A,B,C)$ is tied to the ordered factor positions appearing in the SHG expression.

For the interband term in Eq.~\eqref{eq:chi_e_dyn}, we assign the three interband factors to atom labels and write
\begin{align}
  \chi^{abc}_{e}(-2\omega;\omega,\omega)
  &= \sum_{A,B,C} \chi^{abc}_{e;ABC}(-2\omega;\omega,\omega), \\
  \chi^{abc}_{e;ABC}(-2\omega;\omega,\omega)
  &=
  \frac{e^3}{\hbar^2}
  \int \frac{d\mathbf{k}}{(2\pi)^3}\sum_{nml}
  \nonumber\\
  &\hspace{-0.8em}\times
  \frac{ r^a_{nm,A}\,\{ r^b_{ml,B} r^c_{ln,C} \} }{\omega_{ln}-\omega_{ml}}
  \nonumber\\
  &\hspace{-0.8em}\times
  \left[
      \frac{2f_{nm}}{\omega_{mn}-2\omega}
    + \frac{f_{ln}}{\omega_{ln}-\omega}
    + \frac{f_{ml}}{\omega_{ml}-\omega}
  \right].
  \label{eq:chi_e_ABC}
\end{align}
Here
\begin{align}
  \{ r^b_{ml,B} r^c_{ln,C} \}
  = \frac{1}{2}\Bigl(r^b_{ml,B}\,r^c_{ln,C}+r^c_{ml,C}\,r^b_{ln,B}\Bigr).
\end{align}
In this convention, each atom label follows the Cartesian slot of the factor in which it appears.
Thus, $B$ is attached to the $b$-directed Berry connection and $C$ to the $c$-directed Berry connection.

For the mixed term $\chi^{abc}_i$, we substitute $r^a_{nm}=\sum_A r^a_{nm,A}$ and $\Delta^a_{mn}=\sum_A \Delta^a_{mn,A}$ throughout Eq.~\eqref{eq:chi_i_dyn}.
We then introduce an atom-pair-resolved generalized derivative by applying the sum rule in Eq.~\eqref{eq:gen_deriv_sum_rule} to the decomposed quantities.
To match the index ordering used in our implementation, we define
\begin{align}
  \mathcal{R}^{ab}_{nm,AB}\equiv r^{b}_{nm;a,AB},
\end{align}
where the first superscript denotes the derivative direction and the second denotes the component index.
Accordingly, the first atom label is associated with the derivative direction and the second with the component index.
For a local Hamiltonian,
\begin{align}
  \mathcal{R}^{ab}_{nm,AB}
  &=
  \frac{ r^{a}_{nm,A}\,\Delta^{b}_{mn,B}
        + r^{b}_{nm,B}\,\Delta^{a}_{mn,A} }{\omega_{nm}}
  \nonumber \\
  &\quad+
  \frac{i}{\omega_{nm}}
    \sum_{l}
    \Bigl(\omega_{lm}\,r^{a}_{nl,A} r^{b}_{lm,B}
  \nonumber \\
  &\qquad\qquad
          - \omega_{nl}\,r^{b}_{nl,B} r^{a}_{lm,A}\Bigr).
  \label{eq:gen_deriv_atom}
\end{align}
This definition satisfies the exact sum rule
\begin{align}
  r^{b}_{nm;a}&=\sum_{A,B} r^{b}_{nm;a,AB},
  \nonumber \\
  \mathcal{R}^{ab}_{nm}&=\sum_{A,B}\mathcal{R}^{ab}_{nm,AB}.
\end{align}

With Eq.~\eqref{eq:gen_deriv_atom}, the mixed susceptibility can be written exactly as
\begin{align}
  \chi^{abc}_{i}(-2\omega;\omega,\omega)
  = \sum_{A,B,C}\chi^{abc}_{i;ABC}(-2\omega;\omega,\omega).
\end{align}
We take
\begin{align}
  \chi^{abc}_{i;ABC}(-2\omega;\omega,\omega)
  &= \frac{i e^{3}}{2\hbar^{2}}
  \int \frac{d \mathbf{k}}{(2\pi)^3}
  \nonumber\\
  &\quad\times
  \sum_{n m} f_{nm}\,
  \mathcal{I}^{abc}_{nm;ABC}(\mathbf{k},\omega),
  \label{eq:chi_i_ABC_def}
\end{align}
with
\begin{widetext}
\begin{align}
  \mathcal{I}^{abc}_{nm;ABC}
  &=
  \frac{2}{\omega_{mn}\bigl(\omega_{mn}-2\omega\bigr)}\,
  r^{a}_{nm,A}\!\left(\mathcal{R}^{bc}_{mn,BC}+\mathcal{R}^{cb}_{mn,CB}\right)
  \nonumber\\
  &\quad+\frac{1}{\omega_{mn}\bigl(\omega_{mn}-\omega\bigr)}\,
  \left(\mathcal{R}^{ca}_{nm,CA}\,r^{b}_{mn,B}+\mathcal{R}^{ba}_{nm,BA}\,r^{c}_{mn,C}\right)
  \nonumber\\
  &\quad+\frac{1}{\omega_{mn}^{2}}
  \left(
  \frac{1}{\omega_{mn}-\omega}-\frac{4}{\omega_{mn}-2\omega}
  \right)
  r^{a}_{nm,A}\!\left(r^{b}_{mn,B}\Delta^{c}_{mn,C}+r^{c}_{mn,C}\Delta^{b}_{mn,B}\right)
  \nonumber\\
  &\quad -\frac{1}{2\,\omega_{mn}\bigl(\omega_{mn}-\omega\bigr)}
  \left(r^{b}_{mn,B}\,\mathcal{R}^{ac}_{nm,AC}+r^{c}_{mn,C}\,\mathcal{R}^{ab}_{nm,AB}\right).
  \label{eq:chi_i_ABC}
\end{align}
\end{widetext}
Here the labels $(A,B,C)$ follow the Cartesian slots $(a,b,c)$ of the corresponding factors.
For example, $\mathcal{R}^{bc}_{mn,BC}$ denotes the contribution to $r^{c}_{mn;b}$ with derivative direction $b$ assigned to atom $B$ and component index $c$ assigned to atom $C$.
By construction, summation over $A,B,C$ recovers Eq.~\eqref{eq:chi_i_dyn} exactly for local Hamiltonians.

For the compact static scheme-N and scheme-L formulas, the response is represented by three-band momentum products after the static-limit rearrangement.
The atom-resolved momentum decomposition therefore yields an exact pathway expansion for both schemes.
We retain the same permutation operator $P(abc)$ as in the total expressions.
Here $P(abc)$ acts only on the Cartesian indices $a$, $b$, and $c$; the atom labels $A$, $B$, and $C$ remain attached to the three momentum factors and are not permuted independently.
Accordingly, the same atom label can appear with different Cartesian components in different terms generated by $P(abc)$.

In scheme-N, we write
\begin{align}
  \chi^{abc}_{\mathrm{(N)}}
  =
  \sum_{A,B,C}\chi^{abc}_{\mathrm{(N)};ABC},
\end{align}
with
\begin{widetext}
\begin{align}
  \chi^{abc}_{\mathrm{(N)};ABC}
  &=
  \frac{e^{3}}{2\hbar^{2}}
  \int \frac{d \mathbf{k}}{(2\pi)^3}
  \sum_{n\in V}\sum_{m\in C}\sum_{l\in V}
  P(abc)\,
  \Im\!\left\{ p^{a}_{nm,A}\,p^{b}_{ml,B}\,p^{c}_{ln,C} \right\}
  \frac{1}{S_{mn}^2 \omega_{nm} \omega_{lm}}
  \left( \frac{1}{S_{lm}}+\frac{2}{\omega_{nm}} \right)
  \nonumber\\
  &\quad+
  \frac{e^{3}}{2\hbar^{2}}
  \int \frac{d \mathbf{k}}{(2\pi)^3}
  \sum_{n\in V}\sum_{m\in C}\sum_{l\in C}
  P(abc)\,
  \Im\!\left\{ p^{a}_{nm,A}\,p^{b}_{ml,B}\,p^{c}_{ln,C} \right\}
  \frac{1}{S_{mn}^2 \omega_{mn} \omega_{ln}}
  \left( \frac{1}{S_{ln}}+\frac{2}{\omega_{mn}} \right),
  \label{eq:schemeN_static_ABC}
\end{align}
\end{widetext}
where
\begin{align}
  S_{nm}=\omega_{nm}+f_{nm}\Delta,
  \qquad \Delta>0.
\end{align}
Because $p^{a}_{nm}=\sum_A p^{a}_{nm,A}$, summation over $A,B,C$ recovers the total scheme-N expression exactly.

In scheme-L, we write similarly
\begin{align}
  \chi^{abc}_{\mathrm{(L)}}
  =
  \sum_{A,B,C}\chi^{abc}_{\mathrm{(L)};ABC},
\end{align}
with
\begin{widetext}
\begin{align}
  \chi^{abc}_{\mathrm{(L)};ABC}
  &=
  \frac{e^{3}}{2\hbar^{2}}
  \int \frac{d \mathbf{k}}{(2\pi)^3}
  \sum_{n\in V}\sum_{m\in C}\sum_{l\in V}
  P(abc)\,
  \Im\!\left\{ \tilde{p}^{a}_{nm,A}\,\tilde{p}^{b}_{ml,B}\,\tilde{p}^{c}_{ln,C} \right\}
  \left(
      \frac{1}{S_{nm}^3 S_{lm}^2}
    + \frac{2}{S_{nm}^4 S_{lm}}
  \right)
  \nonumber\\
  &\quad+
  \frac{e^{3}}{2\hbar^{2}}
  \int \frac{d \mathbf{k}}{(2\pi)^3}
  \sum_{n\in V}\sum_{m\in C}\sum_{l\in C}
  P(abc)\,
  \Im\!\left\{ \tilde{p}^{a}_{nm,A}\,\tilde{p}^{b}_{ml,B}\,\tilde{p}^{c}_{ln,C} \right\}
  \left(
      \frac{1}{S_{mn}^3 S_{ln}^2}
    + \frac{2}{S_{mn}^4 S_{ln}}
  \right),
  \label{eq:schemeL_static_ABC}
\end{align}
\end{widetext}
where the scissor-renormalized atom-resolved momentum matrix elements are
\begin{align}
  \tilde{p}^{a}_{nm,A}
  =
  \left[1+\frac{\Delta}{\omega_{nm}}\left(\delta_{nC}-\delta_{mC}\right)\right]p^{a}_{nm,A}.
\end{align}
Since the prefactor depends only on the band pair $(n,m)$, exact additivity is preserved,
\begin{align}
  \tilde{p}^{a}_{nm}=\sum_A \tilde{p}^{a}_{nm,A},
\end{align}
and summation over $A,B,C$ recovers the total scheme-L expression exactly.

\subsection{Unordered atom-triplet-resolved SHG coefficients}
\label{subsec:unordered_atomic_shg}

The ordered contributions introduced above are exact, but their labels depend on the convention used to assign atom indices to the factor positions of a given term.
For physical interpretation, however, one is often interested only in which atoms participate in a given SHG pathway, rather than in the order in which they appear in the formal expression.
This motivates an unordered decomposition.

For any ordered contribution $\chi^{abc}_{X;ABC}$, where $X\in\{e,i,\mathrm{(N)},\mathrm{(L)}\}$ denotes the chosen SHG expression, we define the corresponding unordered atom-triplet contribution by collecting all distinct reorderings of the same atom multiset:
\begin{align}
    \chi^{abc}_{X;\{A,B,C\}}
  \equiv
  \sum_{(A',B',C')\in \mathcal{P}(A,B,C)}
  \chi^{abc}_{X;A'B'C'}.
  \label{eq:unordered_atom_triplet}
\end{align}
Here $\mathcal{P}(A,B,C)$ denotes the set of distinct permutations of the triplet $(A,B,C)$.
If $A$, $B$, and $C$ are all different, $\mathcal{P}(A,B,C)$ contains six elements.
If two labels are identical, it contains three elements.
If $A=B=C$, it contains a single element.
Thus, Eq.~\eqref{eq:unordered_atom_triplet} sums each distinct ordered pathway exactly once and avoids overcounting for repeated labels.

By construction, the unordered atom-triplet contributions satisfy the exact reconstruction formula
\begin{align}
  \chi^{abc}_{X}
  =
  \sum_{\{A,B,C\}} \chi^{abc}_{X;\{A,B,C\}},
  \label{eq:unordered_atom_sumrule}
\end{align}
where the sum runs over all distinct unordered atom triplets.
Equation~\eqref{eq:unordered_atom_sumrule} follows directly from the partition of the full ordered sum into disjoint permutation classes.

The unordered representation is advantageous because it removes the bookkeeping dependence associated with the ordered factor positions and exposes the spatial character of the SHG pathways more directly.
In particular, triplets of the form $\{A,A,A\}$ correspond to on-site contributions in the present AIM partitioning.
Triplets of the form $\{A,A,B\}$ or $\{A,B,B\}$ involve two atoms and therefore represent two-center pathways, whereas triplets of the form $\{A,B,C\}$ with three distinct labels represent multicenter pathways.
The distinction between $\{A,A,B\}$ and $\{A,B,B\}$ remains useful in practice because the repeated label identifies which atom occupies two of the three matrix-element factors in the underlying ordered decomposition.

\subsection{Unordered motif-triplet-resolved SHG coefficients}
\label{subsec:motif_shg}

The unordered atom-triplet decomposition can be coarse-grained further to obtain contributions associated with chemically or structurally meaningful motifs.
We require the motifs to form a disjoint and collectively exhaustive partition of the atomic sites, such that every atom belongs to exactly one motif.
Let $\mathcal{M}$, $\mathcal{N}$, and $\mathcal{L}$ denote three motifs, each defined as a set of atoms.
To preserve consistency with the unordered atom-triplet construction, the motif contribution must be defined by summing over \emph{distinct unordered atom triplets}, rather than by a direct triple sum over atoms.
We therefore define
\begin{align}
  \chi^{abc}_{X;\{\mathcal{M},\mathcal{N},\mathcal{L}\}}
  \equiv
  \sum_{\{A,B,C\}\in \mathcal{U}(\mathcal{M},\mathcal{N},\mathcal{L})}
  \chi^{abc}_{X;\{A,B,C\}},
  \label{eq:motif_def}
\end{align}
where $\mathcal{U}(\mathcal{M},\mathcal{N},\mathcal{L})$ denotes the set of distinct unordered atom triplets $\{A,B,C\}$ whose motif-membership pattern is exactly $\{\mathcal{M},\mathcal{N},\mathcal{L}\}$.
In this way, each unordered atom triplet is counted once, including cases with repeated motifs such as $\{\mathcal{M},\mathcal{M},\mathcal{N}\}$ and $\{\mathcal{M},\mathcal{N},\mathcal{N}\}$.
Equation~\eqref{eq:motif_def} may therefore be viewed as an exact coarse-graining of the unordered atom-triplet decomposition from the atomic level to the motif level.

Because Eq.~\eqref{eq:motif_def} is defined by regrouping the exact unordered atom-triplet contributions, it also satisfies exact additivity:
\begin{align}
  \chi^{abc}_{X}
  =
  \sum_{\{\mathcal{M},\mathcal{N},\mathcal{L}\}}
  \chi^{abc}_{X;\{\mathcal{M},\mathcal{N},\mathcal{L}\}},
  \label{eq:motif_sumrule}
\end{align}
where the sum runs over all distinct unordered motif triplets.
Thus, the motif-resolved decomposition is not an approximation, but an exact regrouping of the underlying atom-resolved pathways.
The resulting motif contributions are exact for the specified crystal and coordinate frame; they are not transferable intrinsic tensors of isolated motifs.

This coarse-grained representation is convenient for interpreting SHG in terms of chemically meaningful building units.
For example, $\{\mathcal{M},\mathcal{M},\mathcal{M}\}$ measures the purely intra-motif contribution, $\{\mathcal{M},\mathcal{M},\mathcal{N}\}$ and $\{\mathcal{M},\mathcal{N},\mathcal{N}\}$ describe two-motif cooperative pathways, and $\{\mathcal{M},\mathcal{N},\mathcal{L}\}$ with three distinct motifs quantifies genuinely nonlocal three-motif cooperation.
This hierarchy provides a within-crystal coarse-grained interpretation of the dominant SHG response channels.
 
    \section{Computational Details}
    \label{sec:details}
    We first use zincblende GaP as a two-atom demonstration of the complete atom-resolved workflow.
All first-principles calculations were carried out with \textsc{GPAW} 25.7.0 in plane-wave mode.\cite{Enkovaara2010}
The ideal primitive cell was constructed using the experimental 300~K lattice constant $a=5.451$~\AA, and no structural relaxation was performed.\cite{Vurgaftman2001}
The GaP calculations employed a 1200~eV plane-wave cutoff, a $\Gamma$-centered $40\times40\times40$ Monkhorst--Pack mesh, and 280 total bands.
Its SHG response was evaluated with the length-gauge scheme-N protocol.\cite{Nastos2005}

We then consider six representative UV and deep-UV nonlinear-optical crystals: the five borates $\beta$-\ce{BaB2O4} (BBO),\cite{Chen1985} \ce{LiB3O5} (LBO),\cite{Chen1989a} \ce{CsB3O5} (CBO),\cite{Wu1993} \ce{CsLiB6O10} (CLBO),\cite{Mori1995} and \ce{KBe2BO3F2} (KBBF),\cite{Mei1995,Wu1996} together with the phosphate \ce{LiCs2PO4} (LCPO).\cite{Li2016c,Shen2016}
Their crystal structures were taken from geometries relaxed with the Perdew--Burke--Ernzerhof (PBE) functional\cite{Perdew1996} and reported in Refs.~\cite{Cheng2018a,Cheng2020b}.
In the conventional description based on anionic groups, BBO contains planar \ce{[B3O6]^3-} rings, while LBO, CBO, and CLBO contain \ce{[B3O7]^5-} units.
KBBF contains planar \ce{[BO3]^3-} triangles within \ce{[Be2BO3F2]} layers, and LCPO contains \ce{[PO4]^3-} tetrahedra.
The plane-wave kinetic-energy cutoff was set to 500~eV for BBO and KBBF, 650~eV for LBO, CBO, and CLBO, and 700~eV for LCPO.
Brillouin-zone integrations employed $\Gamma$-centered Monkhorst--Pack meshes of $6\times6\times6$, $8\times9\times12$, $11\times8\times7$, $9\times9\times9$, $15\times15\times15$, and $7\times7\times4$ for BBO, LBO, CBO, CLBO, KBBF, and LCPO, respectively.

Unless stated otherwise, all remaining parameters were kept at their default \textsc{GPAW} values.
The optical calculations used the same $k$-point meshes as the ground-state calculations.
For all six crystals, the fixed-density optical calculations requested a total number of bands equal to four times the number of occupied bands.
The ten incompletely converged upper bands were excluded from the response sums.

The conventional momentum matrices were evaluated with \textsc{GPAW}.
The atom-resolved matrices were constructed with our in-house \texttt{AIMOPT} Python package from the PAW wave functions, projectors, and numerical settings specified above.
The MBIS,\cite{Verstraelen2016} LISA,\cite{Benda2022,Cheng2025b,Cheng2025c} Hirshfeld,\cite{Hirshfeld1977} Hirshfeld-I,\cite{Bultinck2007} and AVH\cite{Heidar-Zadeh2024} partitioning methods were evaluated with the periodic-grid implementation in \texttt{HORTON-Part} 1.2.0.\cite{HortonPartPeriodic}
The present AVH calculations use the bound-state AVH-B variant.
These five partitioning methods were pointwise normalized and evaluated on the same uniform density grid for each material.
LISA optimizes nonnegative pro-atom coefficients in its element-specific basis, whereas AVH-B expands each pro-atom in the complete set of physically bound spherical charged-atom states.\cite{Heidar-Zadeh2024}

For GaP, all five partitioning methods were carried through the complete atom-resolved analysis.
For the six ultraviolet crystals, MBIS was used for the atom-triplet analysis of every compound.
The sensitivity comparison among the five partitioning methods was performed at the element level for all six crystals by summing the normalized atomic weights of each element.
KBBF was additionally evaluated with all five partitioning methods at the atom level as a representative multiatom showcase.

Throughout this work, we report the SHG response in terms of the Cartesian tensor components $\chi^{abc}$ and the contracted (Voigt) coefficients $d_{ij}$, where $i=1,2,3$ correspond to $a=x,y,z$ and $j=1,\ldots,6$ correspond to the symmetrized index pairs $(bc)=(xx,yy,zz,yz,zx,xy)$.
Within this convention, $d_{ij}=\tfrac{1}{2}\chi^{abc}$, with $a\leftrightarrow i$ and the symmetrized pair $(bc)\leftrightarrow j$.
Because 1064 and 532~nm lie below the absorption edges of the six ultraviolet crystals, the static limit provides a useful approximation to their measured SHG response.
The main material comparison therefore uses static SHG.
We also evaluate BBO explicitly at 1064~nm to verify finite-frequency reconstruction.
 
    \section{Results}
    \label{sec:results}
    \subsection{GaP: a two-atom demonstration}
\label{subsec:gap_demo}

GaP provides the simplest nontrivial example because its primitive cell contains one Ga and one P atom.
Decomposing $\hat{\mathbf p}=\hat{\mathbf p}_{\mathrm{Ga}}+\hat{\mathbf p}_{\mathrm{P}}$ produces four terms in linear response, one for each assignment of the two momentum factors.
In SHG, assigning three factors produces eight ordered terms.
Combining terms with the same set of atomic labels leaves $\{\mathrm{Ga},\mathrm{Ga},\mathrm{Ga}\}$, $\{\mathrm{Ga},\mathrm{Ga},\mathrm{P}\}$, $\{\mathrm{Ga},\mathrm{P},\mathrm{P}\}$, and $\{\mathrm{P},\mathrm{P},\mathrm{P}\}$.
The first and last are on-site contributions and the two mixed groups are two-center contributions; a three-center term cannot occur in a two-atom cell.
Larger cells follow exactly the same expansion and can additionally contain three-center terms.

The calculated PBE indirect gap is 1.62~eV, compared with the measured 300~K value of 2.26~eV.\cite{Panish1969}
The resulting rigid conduction-band shift is approximately 0.64~eV.
The static $\chi^{xyz}$ is 125.55~pm/V without this shift and 84.88~pm/V with the indirect-gap shift.
At 0.117~eV, a long-wavelength experiment gives $\chi^{xyz}=74\pm4$~pm/V, or $d_{14}=37\pm2$~pm/V.\cite{LevineBethea1972}
Earlier static calculations reported 50.3~pm/V\cite{Hughes1996} with FLAPW.
Shoji \textit{et al.} later measured $d_{36}=36.8$~pm/V at a fundamental wavelength of 1.313~$\mu$m, equivalent to $\chi^{xyz}=73.6$~pm/V.\cite{Shoji1997}
Other static calculations reported 75~pm/V\cite{Levine1991} with a scissor-corrected pseudopotential, 83~pm/V\cite{Corso1996} with self-consistent LDA, and 134.9~pm/V\cite{Huang1993} with OLCAO.

For atom $A$, the AIM charge is defined as $q_A=Z_A-\int_{\Omega}w_A(\mathbf r)n(\mathbf r)\,d^3r$.
It is a partition-dependent analysis quantity rather than a unique observable or an exact oxidation state.
For five matched AIM partitions evaluated on the same GaP uniform grid, the absolute two-center fraction spans 80.6--82.2\%, while the complementary on-site fraction spans 17.8--19.4\%.
The common-grid Ga charges are $+0.98\,e$ (MBIS), $+0.29\,e$ (LISA), $+0.12\,e$ (Hirshfeld), $+1.38\,e$ (Hirshfeld-I), and $+0.30\,e$ (AVH-B).
The conventional Hirshfeld charge is small because fixed neutral free-atom pro-densities favor AIM densities close to neutral atoms and therefore suppress charge transfer.\cite{Hirshfeld1977,Bultinck2007}
Hirshfeld-I instead updates the pro-atom populations self-consistently using charged isolated-atom densities, which removes the neutral-reference bias and commonly increases the charge magnitude.\cite{Bultinck2007}
MBIS also adapts its pro-atoms, but uses a minimal Slater-shell representation with variationally optimized populations and radial widths.\cite{Verstraelen2016}
This distinct model gives the intermediate $+0.98\,e$ Ga charge in GaP.
LISA and AVH-B give smaller intermediate values because their pro-atoms are optimized within other constrained reference spaces: an element-specific radial basis for LISA and physically bound charged-atom states for AVH-B.\cite{Benda2022,Cheng2025b,Cheng2025c,Heidar-Zadeh2024}
The spread therefore reflects the different pro-atom models and optimization conditions rather than an error in the density partition.
Every partition reconstructs the same total, with a maximum difference of $1.3\times10^{-12}$~pm/V.
The leading two-center assignment is therefore robust even though the detailed charge and local percentages depend on the AIM definition.
Complete structural and AIM-charge data for GaP are reported in Table~S1 of the Supplemental Material~\cite{SupplementalMaterial}.
\nocite{Chan2024,Tehrani2024,Kim2024,Verstraelen2021,Aspnes1983}

\subsection{Validation of the ultraviolet-crystal calculations}

Table~\ref{tbl:band_gap} compares the calculated PBE gaps with prior calculations and the experimental gaps used for scissors corrections.
The present values differ from the reported VASP values by at most 0.10~eV.
As expected for a semilocal functional, PBE underestimates each experimental gap.
We therefore apply a rigid scissors shift, defined as the difference between the experimental and PBE gaps, to the conduction bands when evaluating the SHG coefficients.

\begin{table*}[!tbp]
  \caption{Calculated, previously reported, and experimental band gaps (eV).
    The experimental values used for the scissors corrections follow Ref.~\cite{Cheng2020b}; representative reported absorption edges are given in parentheses.}
  \centering
  \begin{tabular}{lccc}
    \toprule
    Material & Calc. & Ref. & Exp. \\
    \midrule
    BBO  & 4.79 & 4.80~\cite{Cheng2020b} & 6.72~\cite{Cheng2020b} (185~nm\cite{Chen2009a}, 190~nm\cite{Chen1985}, 193~nm\cite{French1991}, 195~nm\cite{Chen2012-ch3}) \\
    LBO  & 6.37 & 6.38~\cite{Cheng2020b} & 8.28~\cite{Cheng2020b} (155~nm\cite{Chen2012-ch3}, 150~nm\cite{Chen2009a}) \\
    CBO  & 5.44 & 5.34~\cite{Cheng2020b} & 7.44~\cite{Cheng2020b} (167~nm\cite{Kagebayashi1999}) \\
    CLBO & 5.06 & 5.09~\cite{Cheng2020b} & 6.90~\cite{Cheng2020b} (180~nm\cite{Chen2012-ch3}) \\
    KBBF & 6.13 & 6.07~\cite{Cheng2020b} & 8.45~\cite{Cheng2020b} (147~nm\cite{Chen2012-ch3}, 155~nm\cite{Wu1996}) \\
    LCPO & 4.46 & 4.43~\cite{Cheng2018a}  & 7.02~\cite{Cheng2020b} (174~nm\cite{Li2016c}, 190~nm\cite{Shen2016}) \\
    \bottomrule
  \end{tabular}
  \label{tbl:band_gap}
\end{table*}

Table~\ref{tbl:static_shg} compares the static scheme-N coefficients with previous calculations and experiment.
The agreement with prior scheme-N values verifies the implementation.
Relative to scheme-L, the scheme-N values are generally larger in magnitude because the two corrections treat generalized-derivative terms differently.\cite{Nastos2005,Cheng2026}
The largest coefficients of BBO, LBO, CBO, CLBO, and KBBF are broadly consistent with the experimental ranges, although signs and magnitudes vary among reports.
In particular, the calculated KBBF $d_{11}=0.50$~pm/V agrees closely with the experimental value of $0.47\pm0.01$~pm/V.
No direct experimental benchmarks are available for LCPO or several minor tensor components.
Overall, the calculations reproduce established SHG magnitudes sufficiently well to support the atom-resolved analysis below.

\begin{table*}[!tbp]
\small
\renewcommand{\arraystretch}{1.0}
\caption{
  Computed and experimental SHG coefficients $d_{ij}$ at $\omega = 0$ for BBO, LBO, CBO, CLBO, KBBF, and LCPO.
All SHG coefficients are given in pm/V.
The ``Calc.'' column lists the present results obtained with scheme-N and evaluated in the static limit.
The ``Ref.~\cite{Cheng2026}'' column gives static-limit ($\omega = 0$) values obtained with scheme-N without explicitly enforcing Kleinman symmetry.
The ``Ref.~\cite{Cheng2020b}'' column gives static-limit ($\omega = 0$) values obtained with scheme-L without explicitly enforcing Kleinman symmetry, as reported in Ref.~\cite{Cheng2020b}, while the ``Exp.'' column lists the experimental SHG coefficients measured at 1064~nm.
}
  \centering
  \begin{tabular}{llcccc}
    \toprule
    Material & $d_{ij}$    & Calc.               & Ref.~\cite{Cheng2026}    & Ref.~\cite{Cheng2020b} & Exp. \\
    \midrule
    BBO      & $d_{22}$    & -2.13 & -2.10          & -1.77             & $\pm 2.3$\cite{Dmitriev1999} ($\pm 1.6$ ($1 \pm 0.05$)\cite{Chen2012-ch3})  \\
             & $d_{31}$    & -0.06 & 0.03          & -0.02             & $\mp 0.16$\cite{Dmitriev1999} $(0.070 \pm 0.03)\times d_{22}$\cite{Chen2012-ch3} \\
             & $d_{33}$    & 0.01 & 0.02          & 0.01              & $0.0$\cite{Chen2012-ch3} \\
             & $d_{15}$    & 0.07 & 0.03          & -0.04             & --  \\
    \midrule
    LBO      & $d_{31}$    & -0.78 & -0.81          & -0.64             & $\mp 0.67$\cite{Dmitriev1999} $\mp 0.98$\cite{Chen2012-ch3}\\
             & $d_{32}$    & 0.91 & 0.93          & 0.75              & $\pm 0.85$\cite{Dmitriev1999} $\mp 1.05$\cite{Chen2012-ch3} \\
             & $d_{33}$    & -0.06 & -0.06          & -0.04             & $\pm0.04$\cite{Dmitriev1999} $\pm 0.06$\cite{Chen2012-ch3} \\
             & $d_{15}$    & -0.81 & -0.83          & -0.69             &  \\
             & $d_{24}$    & 0.92 & 0.89          & 0.76              &  \\
    \midrule
    CBO      & $d_{14}$    & 1.05 & 1.03          & 0.80              & $1.49$\cite{Dmitriev1999} 0.86\cite{Chen2012-ch3}  \\
             & $d_{25}$    & 1.03 & 1.00          & 0.82              &  \\
             & $d_{36}$    & 1.00 & 0.98          & 0.71              &  \\
    \midrule
    CLBO     & $d_{14}$    & -0.89& -0.88          & 0.75              &  \\
             & $d_{36}$    & -0.81& -0.89          & 0.66              & $0.86$\cite{Dmitriev1999} 0.74\cite{Chen2012-ch3}  \\
    \midrule
    KBBF     & $d_{11}$    & 0.50& 0.54          & -0.45             & $0.47\pm0.01$\cite{Chen2012-ch3} \\
             & $d_{14}$    & 0.01& -0.02          & --                 &  \\
    \midrule
    LCPO     & $d_{31}$    & 1.02& 0.88          & 0.68              &  \\
             & $d_{32}$    & -0.49& -0.44          & -0.35             &  \\
             & $d_{33}$    & -0.90& -0.77          & -0.71             &  \\
             & $d_{15}$    & 1.04& 0.91          & 0.79              &  \\
             & $d_{24}$    & -0.44& -0.39          & -0.30             &  \\
    \bottomrule
  \end{tabular}
  \label{tbl:static_shg}
\end{table*}
 
\subsection{AIM charges}

\begin{figure}[!tbp]
  \centering
  \includegraphics[width=0.90\columnwidth]{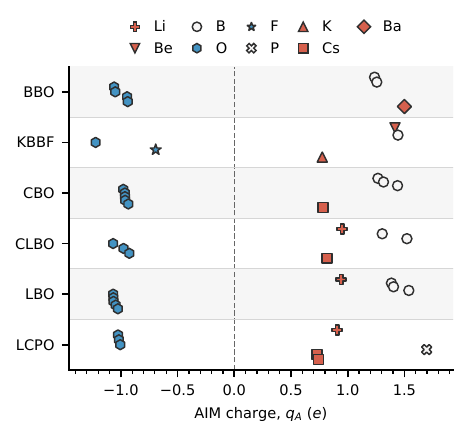}
  \caption{MBIS charges of the symmetry-inequivalent atomic sites in BBO, KBBF, CBO, CLBO, LBO, and LCPO.
    Each marker denotes one representative site and is distinguished by element.}
  \label{fig:aim_unique_charges}
\end{figure}

Figure~\ref{fig:aim_unique_charges} shows the MBIS charges of the symmetry-inequivalent sites in the six crystals.
Alkali and alkaline-earth sites carry positive charges from $+0.73$ to $+1.50\,e$.
The B centers in the borate groups range from $+1.24$ to $+1.54\,e$, while P in LCPO reaches $+1.70\,e$.
Oxygen ranges from $-0.93$ to $-1.22\,e$, whereas F in KBBF is less negative at $-0.69\,e$.
These expected trends distinguish the electropositive sites from the atoms in the anionic groups, while remaining dependent on the chosen AIM partition.
The method-dependent spread follows the pro-atom models and optimization conditions discussed above for GaP.
Values for symmetry-equivalent sites agree within $4.94\times10^{-4}\,e$, showing that the numerical partition respects the crystallographic equivalences.
Complete charge and crystallographic tables are provided in Tables~S2--S7 of the Supplemental Material.
All six tables use the same format and report lattice parameters, fractional coordinates, representative symmetry-unique sites, and method-resolved AIM charges for every atom in the six ultraviolet crystals.

\subsection{Unordered atom-triplet contributions}

\begin{figure}[!tbp]
  \centering
  \includegraphics[width=0.90\columnwidth]{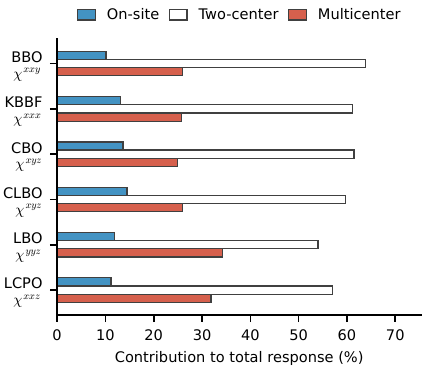}
  \caption{Signed on-site, two-center, and multicenter percentages for the leading scheme-N SHG component of the six crystals.}
  \label{fig:unord_summary}
\end{figure}

We next group the exact unordered atom-triplet decomposition into on-site $\{A,A,A\}$, two-center $\{A,A,B\}/\{A,B,B\}$, and multicenter $\{A,B,C\}$ pathways.
Figure~\ref{fig:unord_summary} summarizes the signed contributions to the leading component of all six crystals.
Two-center pathways contribute 54--64\%, compared with 10--15\% from on-site pathways and 25--34\% from multicenter pathways.
The dominance of the two-center class is therefore systematic, but multicenter coupling remains substantial.
LBO and LCPO lie toward the upper end of the multicenter range, with a corresponding reduction in their two-center fractions.
This trend indicates a stronger role for extended multicenter response pathways in these two compounds.
The absolute decomposition preserves the same hierarchy, showing that the response is not reducible to either isolated on-site terms or a purely local two-center picture.
Table~S8 and Fig.~S1 of the Supplemental Material give the complete signed and absolute class decompositions.

\begin{figure}[!tbp]
  \centering
  \includegraphics[width=\columnwidth]{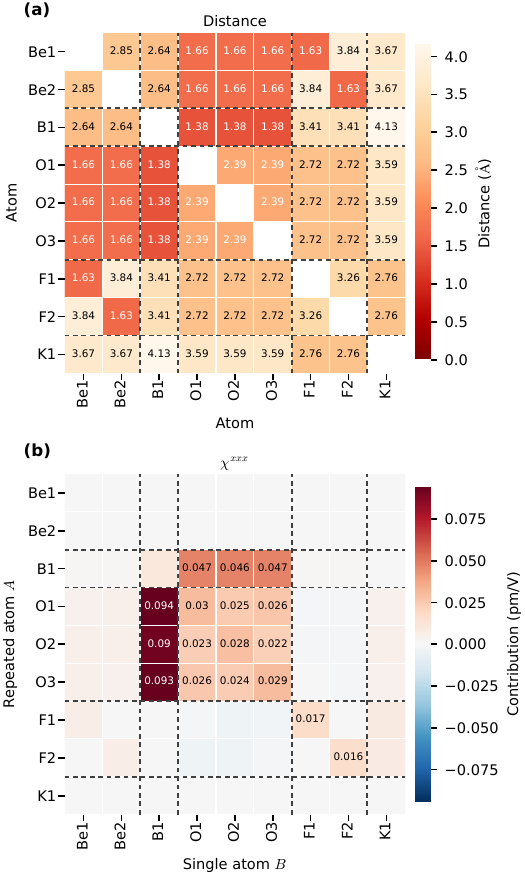}
  \caption{Representative unordered atom-triplet analysis for $\chi^{xxx}$ of KBBF using scheme-N.
    (a) Interatomic distances in \AA.
    (b) On-site (diagonal) and two-center (off-diagonal) contributions in pm/V.
    The dashed lines separate atomic species.}
  \label{fig:unord_kbbf}
\end{figure}

Figure~\ref{fig:unord_kbbf} resolves the leading $\chi^{xxx}$ component of KBBF against the interatomic-distance matrix.
The dominant entries are short-range B--O and O--O pathways associated with the \ce{[BO3]} groups, while K- and F-containing entries are smaller.
For an atom pair $(A,B)$, the combined two-center contribution is the sum of the $\{A,A,B\}$ and $\{A,B,B\}$ pathways.
The ranked pair contributions and distance dependence show that contributions beyond 3~\AA{} are negligible (Figs.~S2 and S3 and Table~S9 of the Supplemental Material).
The leading response of LBO is similarly dominated by short-range B--O and O--O pathways (Figs.~S4--S6 and Table~S10).
BBO instead combines strong B/O terms with appreciable O--Ba pathways, whose largest pair terms occur near 2.9~\AA{} (Figs.~S7--S9 and Table~S11).
In CBO and CLBO, the largest individual two-center terms involve O--Cs, although the summed B/O contribution remains larger than the total Cs-associated contribution (Figs.~S10--S15 and Tables~S12 and S13).
In LCPO, O--Cs gives the largest individual two-center contribution, and its summed contribution exceeds that from O--P (Figs.~S16--S18 and Table~S14).
The decomposition therefore shows that B/O pathways dominate the leading responses of KBBF and LBO.
BBO, CBO, CLBO, and LCPO instead have leading responses with substantial mixed pathways involving polarizable cations.

\subsection{Element-triplet contributions}

\begin{figure}[!tbp]
  \centering
  \includegraphics[width=0.90\columnwidth]{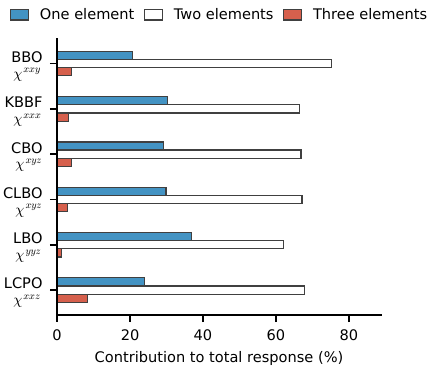}
  \caption{Signed one-, two-, and three-element percentages for the leading scheme-N SHG component of each crystal.}
  \label{fig:elem_class_summary}
\end{figure}

Conventional motif analyses require chemical units to be defined \emph{a priori}, such as \ce{[BO3]} in KBBF and \ce{[B3O6]} in BBO.
To avoid a system-specific motif choice and enable a consistent comparison across materials, we instead group atoms by chemical element.
Figure~\ref{fig:elem_class_summary} groups the leading component into one-, two-, and three-element pathways.
Two-element pathways contribute 62--75\% of the signed response, compared with 21--37\% from one-element pathways.
Three-element pathways contribute only 1--4\% in the borates and 8\% in LCPO.
This hierarchy is consistent with the atom-level result.
The complete signed and absolute class data are reported in Table~S15 and Fig.~S19 of the Supplemental Material.

\begin{figure*}[!tbp]
  \centering
  \includegraphics[width=\textwidth]{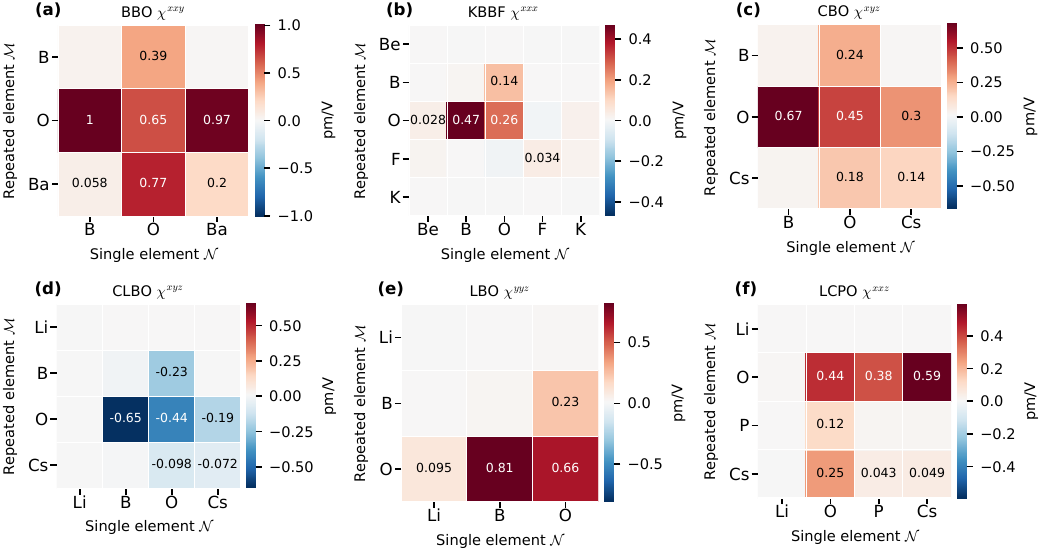}
  \caption{Representative element-triplet contributions to the leading SHG components of BBO, KBBF, CBO, CLBO, LBO, and LCPO.
    Only the leading one- and two-element pathways are shown.
    The repeated element $\mathcal{M}$ appears twice in $\{\mathcal{M},\mathcal{M},\mathcal{N}\}$, and the single element $\mathcal{N}$ appears once.
    Values are in pm/V.}
  \label{fig:elem_triplet_main}
\end{figure*}

The resolved heat maps in Fig.~\ref{fig:elem_triplet_main} reveal two material classes.
B/O-only pathways dominate the leading responses of KBBF and LBO: $\{\mathrm{B,O,O}\}$ is the largest pathway, and the corresponding absolute fractions span 64.2--86.1\% and 93.4--98.7\%, respectively, across five partitions.
In KBBF, Be-, F-, and K-containing triplets provide smaller corrections.
In LBO, Li-containing pathways remain weak.
This assignment agrees with earlier RSAC-based analyses of the borate groups.\cite{Lin2000,Chen2005}
The complete scheme-N data are given in Tables~S17 and S21 and Figs.~S20 and S21 of the Supplemental Material.

BBO, CBO, and CLBO show substantial mixed contributions involving B/O atoms and cations.
Their mixed absolute fractions span 44.5--49.9\%, 27.2--30.6\%, and 18.8--23.6\%, respectively.
In BBO, $\{\mathrm{B,O,O}\}$ contributes 1.00~pm/V to $\chi^{xxy}$, while the oxygen-mediated $\{\mathrm{Ba,O,O}\}$ and $\{\mathrm{Ba,Ba,O}\}$ pathways contribute 0.97 and 0.77~pm/V.
The Ba-only term is much smaller, showing that the cation effect is primarily cooperative rather than an isolated on-site response.
CBO and CLBO display the analogous pattern through oxygen-mediated Cs pathways, and their purely Cs on-site channels are non-negligible.
This interpretation agrees qualitatively with earlier RSAC analyses.\cite{Lin1999,Lin2001}
Quantitatively, the Ba-only on-site term contributes approximately 5\% in BBO, compared with approximately 16\% in the RSAC analysis.\cite{Lin1999}
RSAC assigns part of the off-site Ba--O response to a prescribed Ba-centered sphere.
The present AIM decomposition instead resolves this response into explicit Ba--O pathways using spatial weight functions.
The different region definitions also explain the analogous difference for Cs in CBO and CLBO.
The corresponding complete scheme-N data are reported in Tables~S25, S29, and S33 and Figs.~S22--S24 of the Supplemental Material.

LCPO has the strongest mixed balance: pathways involving both P/O and cations account for 44.8--59.6\% of the absolute contribution.
Its leading $\chi^{xxz}$ pathways include $\{\mathrm{Cs,O,O}\}$, $\{\mathrm{O,O,O}\}$, and $\{\mathrm{P,O,O}\}$, whereas Li-containing pathways are comparatively small.
The negative $\chi^{yyz}$, $\chi^{zyy}$, and $\chi^{zzz}$ components are controlled by the same pathway classes with opposite signs.
For the leading component, the on-site hierarchy $\{\mathrm{O,O,O}\}>\{\mathrm{Cs,Cs,Cs}\}>\{\mathrm{P,P,P}\}>\{\mathrm{Li,Li,Li}\}$ agrees with the trend from ART,\cite{Cheng2018a} apart from the reversed Li/P ordering reported there.
The O--Cs off-site contribution is nevertheless larger than every on-site term.
Across all five partitions, the stable conclusion is therefore whether B/O-only or P/O-only pathways dominate the leading response or whether mixed pathways involving cations are substantial.
Exact pathway ranks and local percentages remain partition dependent.
The complete LCPO scheme-N data are given in Table~S37 and Fig.~S25 of the Supplemental Material.

\subsection{Partition robustness and numerical convergence}

We finally assess how the preceding atom- and element-triplet assignments depend on the AIM partition and numerical convergence.
For five matched KBBF partitions, the on-site, two-center, and multicenter ranges are 17.7--41.1\%, 56.8--68.7\%, and 2.1--13.8\%, respectively.
Every partition sum is exact within numerical precision.
Thus, the leading two-center assignment is robust, whereas detailed local percentages and charges are analysis dependent.

\begin{figure*}[!t]
  \centering
  \includegraphics{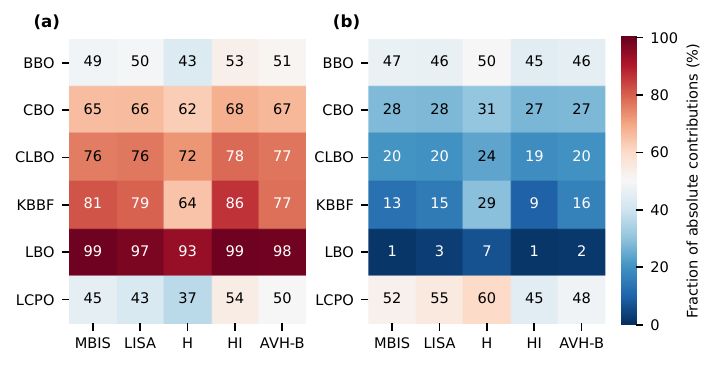}
  \caption{Absolute leading-component fractions obtained by grouping the atomic weights from five AIM partitioning methods by element.
    Within each material, all methods use the same uniform grid.
    Panel (a) contains contributions whose three element indices are all B/O in the borates or all P/O in LCPO.
    This element-based classification does not isolate a geometrical anionic group.
    Panel (b) contains mixed contributions with at least one B/O or P/O element and one other element.
    Contributions containing only the remaining elements are not shown; their complete ranges are reported in Table~S40 of the Supplemental Material.
    H denotes Hirshfeld, HI denotes Hirshfeld-I, and AVH-B denotes the bound-state additive variational Hirshfeld model.
    LISA denotes the linear approximation of iterative stockholder analysis.}
  \label{fig:multimethod_partition_robustness}
\end{figure*}

For the six ultraviolet crystals, we apply the same five partitioning methods at the atom level and then sum the normalized atomic weights by element.
Figure~\ref{fig:multimethod_partition_robustness} shows the B/O-only or P/O-only fractions and the fractions from mixed pathways containing other elements for their leading components.
These are element-based classes and do not represent the response of a geometrically isolated anionic group.
Fractions from pathways containing only the remaining elements are reported in Table~S40.
Table~S40 and Fig.~S26 of the Supplemental Material also provide an independent implementation cross-check: the reconstructed totals differ by at most $7.9\times10^{-12}$~pm/V.

Conventional Hirshfeld is the principal quantitative outlier among the five stockholder partitions.
Its fixed neutral pro-atoms do not adapt to charge transfer in the chemical environment.\cite{Hirshfeld1977,Bultinck2007}
For KBBF, Hirshfeld assigns 64.2\% to B/O-only pathways and 29.1\% to mixed pathways.
The four adaptive methods (MBIS, LISA, Hirshfeld-I, and AVH-B) instead span 77.4--86.1\% and 9.5--16.2\%, respectively.
We therefore regard conventional Hirshfeld as a systematic fixed-neutral-reference exception, rather than as a numerical failure.
The adaptive methods still exchange closely competing pathway ranks in BBO and LCPO.
Thus, the broad material classification is more robust than an exact pathway ordering.

The BBO calculation uses a 1.92~eV scissors shift and retains 470 total bands (120 occupied and 350 unoccupied).
A separate cutoff test shows that the interband and mixed interband--intraband terms stabilize from 800 bands onward, while their residual errors largely cancel in the total response.
Within the common 1000-band cache, the native-frame totals evaluated with 470 and 1000 bands differ by less than 0.001~pm/V.
The motif sum reconstructs an independent total with a relative maximum error of $3.8\times10^{-8}$.
At 1064~nm, independently contracted pathways reconstruct the total with the complex relative error $|\chi_{\mathrm{sum}}-\chi_{\mathrm{total}}|/|\chi_{\mathrm{total}}| =1.98\times10^{-7}$.
Table~S42 and Figs.~S27(a) and S27(b) of the Supplemental Material report the convergence and finite-frequency reconstruction data.
Tables~S43 and S44 compare the measured computational costs of the electronic-structure, matrix-element, SHG, and AIM-partitioning steps, showing that AIM weight construction is not the computational bottleneck.
 
\begingroup
    \let\FloatBarrier\relax
    \section{Summary}
    \label{sec:summary}
    \endgroup
    We developed an exactly additive framework that connects atom-resolved PAW momentum matrices to macroscopic optical response.
Two matrix factors give atom-pair linear susceptibility, while three give the atom-resolved second-order response used here for SHG.
The same construction formally extends to third order without redefining the atomic electronic quantities.
Across the six crystals, two-center channels dominate the principal response.
On-site channels are smaller, while multicenter channels provide a substantial secondary contribution.
Grouping the atom-resolved pathways by element shows that B/O-only pathways dominate the leading responses of KBBF and especially LBO.
BBO and LCPO have the largest mixed fractions involving cations, while CBO and CLBO are intermediate.
This interpretation persists across five AIM partitioning methods evaluated at the atomic level on a common material-specific grid.
Conventional Hirshfeld is the principal quantitative outlier because it retains fixed neutral pro-atoms, whereas the four adaptive stockholder methods cluster more closely.
Closely competing pathways can nevertheless exchange rank within the adaptive subset.

The covalent GaP and ionic KBBF benchmarks are likewise dominated by two-center channels across five matched AIM partitions.
Every matched test reproduces the same total response but shifts detailed percentages.
The decomposition framework is applicable across distinct bonding regimes and AIM definitions, whereas the assigned local values remain partition-dependent analysis quantities rather than unique observables.
Separate BBO band-convergence tests quantify cancellation in the total, and the calculation at 1064~nm verifies finite-frequency additivity.
Together, these results establish a reproducible route for identifying local response channels while preserving the bulk nonlinear response exactly.
 
    \section*{Data Availability}
    The data supporting the findings of this study are available from the corresponding authors upon reasonable request.
The \texttt{AIMOPT} software used for the atom-resolved analysis is under development and is not yet publicly available.
 
    \begin{acknowledgments}
        This work was supported by the Key Research Program of Frontier Sciences, National Natural Science Foundation of China (T2600851, 22193044, 52403305, 22361132544), CAS Project for Young Scientists in Basic Research (YSBR-024), the Strategic Priority Research Program of the Chinese Academy of Sciences (XDB0880000).
     \end{acknowledgments}

\makeatletter
    \immediate\write\@auxout{\string\citation{apsrev42Control}}
    \makeatother
    \bibliography{references}

\end{document}